\documentclass[aps,pra,showpacs,twocolumn,letterpaper,nobibnotes,superscriptaddress]{revtex4-1}
\usepackage{amssymb}
\usepackage{amsmath}
\usepackage{graphicx}
\usepackage{mathrsfs}
\usepackage{url}
\usepackage{color}

\begin{document}
\title{Strictly nonclassical behavior of a mesoscopic system}
\author{Jiazhong Hu}
\author{Zachary Vendeiro}
\author{Wenlan Chen}
\author{Hao Zhang}
\author{Robert McConnell}
\affiliation{Department of Physics and Research Laboratory of Electronics, Massachusetts Institute of Technology, Cambridge, Massachusetts 02139, USA}
\author{Anders S. S\o{}rensen}
\affiliation{The Niels Bohr Institute, University of Copenhagen, Blegdamsvej 17, DK-2100, Copenhagen, Denmark}
\author{Vladan Vuleti\'c}
\affiliation{Department of Physics and Research Laboratory of Electronics, Massachusetts Institute of Technology, Cambridge, Massachusetts 02139, USA}

\begin{abstract}
We experimentally demonstrate the strictly nonclassical behavior in a many-atom system using a recently derived criterion [E. Kot et al., \textit{Phys. Rev. Lett.} \textbf{108}, 233601 (2012)] that explicitly does not make use of quantum mechanics. We {\color{black}t}hereby show that the magnetic moment distribution measured by McConnell et al. [R. McConnell et al., \textit{Nature} \textbf{519}, 439 (2015)] in a system with a total mass of $2.6\times 10^5$ atomic mass units is inconsistent with classical physics. Notably, the strictly nonclassical behavior affects an area in phase space $10^3$ times larger than the Planck quantum $\hbar$.
\end{abstract}
\maketitle

Ever since the advent of modern quantum theory almost a century ago, one perplexing question has been the boundary between quantum mechanics and classical physics. Considering just two particles, Bell showed \cite{bell,CHSH} that reasonable assumptions consistent with classical physics lead to predictions that are inconsistent with the measurement results \cite{aspect1,aspect2,loop1,loop2,Schmied441}. For macroscopic systems, Schr\"odinger's gedanken experiment of a simultaneously dead and alive cat highlights the question about the quantum-classical boundary in larger and larger systems \cite{GHZ,Leibfried04062004,Roos04062004,PhysRevLett.106.130506,Size4,Size1,Size2,Size3}. Within the conceptual framework of quantum mechanics, one can establish definite and quantitative criteria for entanglement \cite{WF,Terhal2000319,PhysRevLett.92.087902,PhysRevLett.84.2726,PhysRevLett.84.2722}, and hence potential non-classicality. {\color{black}However, one may argue that an experiment never confirms a theory as valid, but only fails to violate it.} Therefore, mere consistency with results derived from quantum mechanics, does not rule out a description within classical physics. It is therefore {\color{black}interesting} to identify and experimentally test criteri{\color{black}a} that are formulated within classical physics, without assuming the validity or concepts of quantum mechanics \cite{PhysRevLett.105.010501}.

\begin{figure*}
\begin{center}
  \includegraphics[width=5.2in]{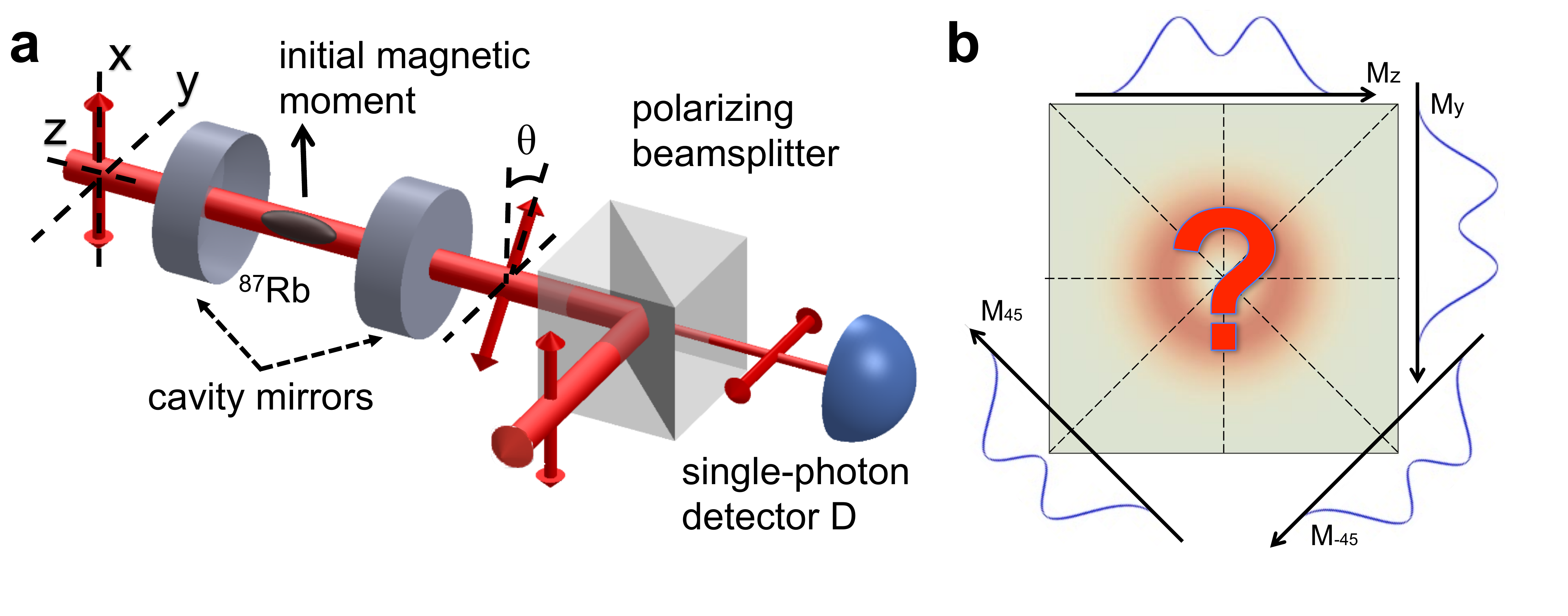}\\
  \caption{\textbf{a,} Experimental setup. $N=3\times 10^3$ $^{87}$Rb atoms are trapped in an optical cavity by a far detuned dipole trap, and prepared in the state 5S$_{1/2}$, F=1. The magnetic moment of the atomic ensemble $M=N\mu_B$ is initialized along the $\hat x$ axis, perpendicular to the cavity axis. A vertically polarized weak incident light pulse experiences a weak Faraday polarization rotation due to the atomic magnetization such that a horizontally polarized photon is sometimes detected. If detector $D$ measures such a photon, the magnetization of the Rb gas is measured along $\hat y$, $\hat z$, or $ {1\over \sqrt{2}}(\hat y \pm \hat z)$, yielding magnetization distributions schematically indicated in (b). In classical physics, the measured distributions should arise from an underlying probability distribution $\rho (M_y,M_z)$. Here, using a variation of the criterion  in Ref. \cite{PhysRevLett.108.233601}, we show that there is no classical probability distribution $\rho (M_y,M_z)$ consistent with the measured result, thereby ruling out any classical description of the magnetic moment.}
  \end{center}
\end{figure*}

{\color{black}In quantum optics, criteria that distinguish nonclassical states of light from classical ones \cite{anti2,anti3,Klyshko1,Klyshko2,anti2,Sta1,Sta2,Sta3,Sta4} were developed early, and tested successfully in experiments \cite{anti1,Sta5}, such as antibunching \cite{anti1,anti2,anti3}, Klyshko's criterion \cite{Klyshko1,Klyshko2}, and nonclassical statistical properties \cite{anti2,Sta1,Sta2,Sta3,Sta4,Sta5}. Some of these demonstrations \cite{anti1,Sta5} have become standard methods to verify classes of nonclassical states, such as the single-photon states \cite{anti1}, and sub-Poissionian states \cite{Sta5}.}

{\color{black}More tests have been performed on atomic and molecular systems.} One such approach is to test the wave properties of larger and larger molecules, and one day perhaps even living objects, through double-slit interference experiments \cite{RMPZ}. The largest objects to date for which matter-wave interference fringes have been observed, are molecules consisting of 430 atoms, with a total mass of 6910 atomic mass units (amu) \cite{Size4}, and interference experiments with even larger molecules appear possible \cite{Size1,Size2,Size3}. Another possible test ground are superconducting qubits, where superpositions of current involving several thousand electrons have been inferred \cite{Knee}.

Other methods to detect the breakdown of classical physics have been proposed \cite{PhysRevLett.84.1849,PhysRevA.65.033830,PhysRevA.83.052113}, and some of them have been verified experimentally by measurements on weak light fields \cite{PhysRevA.65.033830,PhysRevA.86.042108}. Recently, Kot et al. \cite{PhysRevLett.108.233601} {\color{black}have} extended the results of Ref. \cite{PhysRevA.83.052113} to show that certain phase space distributions are inconsistent with classical physics, and have derived a quantitative formalism to identify such distributions without involving quantum mechanical concepts. We term this the phase space distribution (PSD) criterion. The PSD criterion has already been experimentally verified for the electromagnetic field associated with single photon \cite{PhysRevLett.108.233601}, and expanded theoretically to some more general cases \cite{generaltest} in quantum optics.

In this Rapid Communication, we apply a modified PSD criterion to the magnetization measurement of the atomic ensemble reported in Ref. \cite{NaMC}. We verify the breakdown of any description of the system in terms of a classical distribution of the magnetic-moment with 98\% confidence. This shows that a system with total mass $M=2.6\times 10^5$ amu and a size of a few millimeters can be manifestly non-classical, further extending the non-classical domain to more massive systems without involving quantum mechanical concepts in the analysis. Notably, unlike the single-photon case \cite{PhysRevLett.108.233601}, where the breakdown of the classical description is associated with small structures in phase space of area $\sim\hbar$, in our $N$-atom system the classical description is violated by features in the phase space distribution function of size $\sim N\hbar\sim 10^3 \hbar$, {\color{black}which contains $10^3$ allowed states \footnote{Note that the PSD criterion requires no knowledge of the total spin, and therefore many different quantum states are potentially consistent with the measured phase space distribution.}}.

The experiment reported in Ref. \cite{NaMC} uses a cold gas of $N=3\times 10^3$ rubidium 87 atoms trapped in an optical cavity to generate the phase space distribution of interest. The atoms are prepared in the $5S_{1/2},F=1$ ground-state hyperfine manifold where each atom has a magnetic moment of one Bohr magneton $\mu_B$. All atoms are initialized with their magnetic moment pointing along the $\hat x$ axis, perpendicular to the cavity axis. Weak probe light, linearly polarized along $\hat x$, is incident onto the cavity. It is resonant with a cavity mode and detuned $\Delta/(2\pi)=-200$~MHz from the $^{87}$Rb D$_2$ hyperfine transition F=1 to F$^\prime$=0. The incident light experiences a weak Faraday polarization rotation due to magnetization fluctuations of the atomic ensemble. In about 5\% of the cases, a photon emerges with a polarization orthogonal to the incident polarization and is detected on detector $D$ (Fig.~1); subsequently the magnetic moment of the ensemble $\vec M$ is measured with a stronger pulse. We consider only those magnetizations of the atomic ensemble where the detector $D$ has registered a photon, and show that the associated magnetization distribution for this set of ensembles violates the PSD criterion. {\color{black}We emphasize that while the preparation process, {\color{black}using particular atomic states, is} rooted in quantum mechanics, the subsequent classical analysis performed here merely considers an ensemble of prepared magnetizations, and does not rely on the specifics of the preparation procedure.}

In order to observe the magnetic moment distribution along different axes, the magnetic moment of each atom in the ensemble is rotated by an angle $\beta=0,\pi/4,\pi/2,3\pi/4$ along the $\hat x$ axis before measuring the squared magnetic moment $M^2_z$. Thus $\beta=0$ corresponds to measuring $M^2_z$, and $\beta=\pi/2$ corresponds to measuring $M^2_y$. We combine all the data for different angles $\beta$ to obtain the rotationally averaged distribution in the $y-z$ plane. This eliminates all structures due to higher-order moments that are not rotationally invariant.

\begin{figure}
  \includegraphics[width=2.5in]{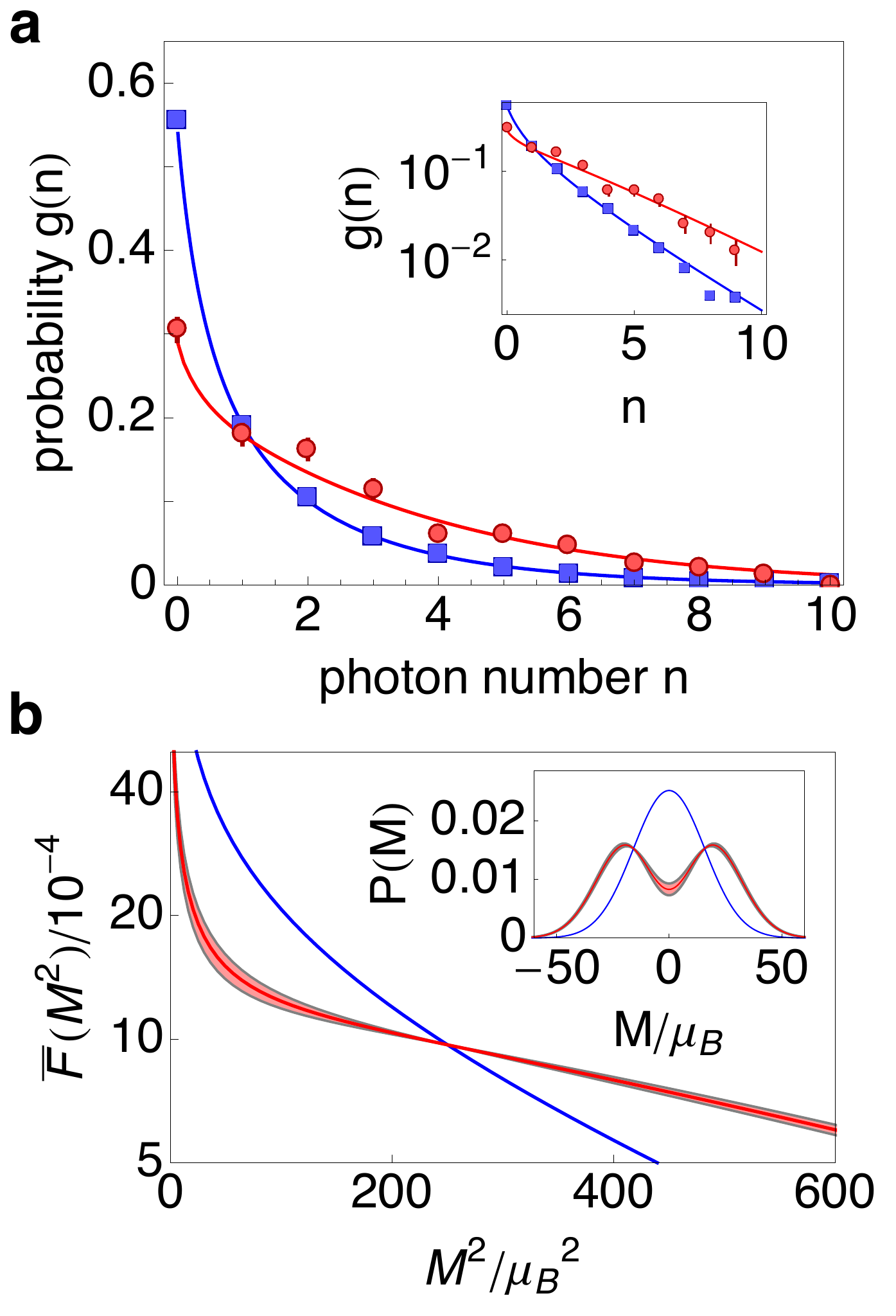}\\
  \caption{Reconstructing the rotationally averaged marginal atomic magnetic moment distribution $\bar F(M^2)$ from the measured photon number distribution $g(n)$. \textbf{a} shows $g(n)$ (red solid circles) for the distribution of interest after registering a {\color{black}heralding click} on detector $D$. The blue squares show the distribution for the initial state with the atomic magnetic-moments prepared along the $\hat x$ axis for reference. For the red solid line, we first apply Eq.~\ref{eq3} solving for the magnetic moment distribution $\bar F(M^2)$, and then convert the distribution back into expected photon counts. This tests the numerical reliability of our data processing method. The inset shows a log-linear scale plot of the same data, which displays the large-photon-number behavior more clearly. \textbf{b}, The reconstructed magnetic-moment distribution $\bar F(M^2)$ using our method. The blue line is for the reference state. The inset shows for illustration purposes the rotationally averaged distribution $P(M)$ assuming $\langle M\rangle=0$. All the error bars and error bands represent one standard deviation.}
\end{figure}

The measurement in Ref. \cite{NaMC} is achieved by sending a stronger light pulse {\color{black}and measuring its Faraday polarization rotation due to the atomic magnetization. By measuring the light emerging with orthogonal polarization compared to the input polarization (along $\hat x$), we can then determine} the square of the magnetization $M^2_z$. For a given magnetization $M_z$, the ensemble rotates the light polarization by a small angle $\theta=\phi M_z$, where $\phi=0.0012/\mu_B$ and $\mu_B$ is the Bohr magneton \cite{PhysRevA.88.063802}. The mean photon number registered on the detector $D$ is then $\chi(M^2_z)=q n_{in}(\phi M_z)^2$, where $q=0.3$ is the overall detection efficiency, {\color{black}and $n_{in}=2\times 10^4$ is the average number of photons in the measurement pulse}. For a given $M_z$, individual photons in the measurement pulse are transmitted independently from one another; therefore for $n_{in}$ input photons the probability to detect exactly $n$ photons is given by the Poisson distribution
\begin{equation}
p(n,M^2_z)=e^{-\chi(M^2_z)}{[\chi(M^2_z)]^n\over n!}.
\end{equation}
For any chosen measurement angle $\theta$ the detected photon distribution $g_{\theta}(n)$ is related to the underlying magnetic-moment distribution $F_{\theta}(M^2_\theta)$ by
\begin{equation}
g_{\theta}(n)=\int d(M^2_\theta) F_{\theta}(M^2_\theta)p(n,M^2_\theta),
\end{equation}
and the angle-averaged measured photon number distribution $g(n)=(2\pi)^{-1}\int d\theta g_\theta(n)$ is related to the angle-averaged magnetic-moment distribution $\bar F(M^2)=(2\pi)^{-1}\int d\theta F_\theta(M^2_\theta=M^2)$ by $g(n)=\int d(M^2)\bar F(M^2)p(n,M^2)$.

To find $\tilde F(M^2)$ from $g(n)$, we introduce a new function 
{\color{black}\begin{equation}
G(\tilde M)=\sum_n g(n)e^{-qn_{in}(\phi\tilde M)^2}{n!\over (2n)!}[4qn_{in}(\phi\tilde M)^2]^n.\label{eq3}
\end{equation}}
{\color{black}It can be shown that $G(\tilde M)$ equals} the convolution of $M\bar F(M^2)$ with the function $f(M)=e^{-qn_{in} \phi^2M^2}$, {\color{black}$G(\tilde M)=\int_{-\infty}^{+\infty} d(M)M e^{-qn_{in} \phi^2(M-\tilde M)^2} \bar F(M^2)$}. The Fourier transform of a convolution equals the product of the Fourier transform of the two individual functions, i.e. {\color{black}$\mathcal S_G(\omega)=\mathcal S_{M\bar F}(\omega)\times \mathcal S_f(\omega)$}, where $\mathcal S_G$ denotes the Fourier transform of the function $G$, and $\omega$ is {\color{black}the} variable after the Fourier transformation. 

We find $G(\tilde M)$ from the measured photon number distribution $g(n)$ according to Eq. (\ref{eq3}), Fourier transform it, and apply the inverse Fourier transform to $S_G(\omega)/S_{M\bar F}(\omega)=S_f(\omega)$ to find the underlying magnetic moment distribution $\bar F(M^2)$ (Fig.~2). In this process, only the Poissonian character of the detected photon number distribution for a given $M^2$ is used to reconstruct $\bar F(M^2)$.

To show that the obtained distribution $\bar F(M^2)$ cannot be obtained from classical physics, we follow the procedure for the PSD criterion \cite{PhysRevLett.108.233601}. We define $M_\rho=\sqrt{M^2_z+M^2_y}$ for convenience and calculate the mean value $\langle\mathscr F\rangle$ for a {\color{black}non-negative} trial function, defined as
\begin{equation}
\mathscr F(M_\rho)=\left(1+\sum_{k=1}^{N_c/2}C_{2k}M^{2k}_\rho\right)^2
\end{equation}
for a given magnetization distribution $\bar F(M^2)$, where the coefficients are chosen according to the relation
\begin{equation}
\sum_{l=1}^{N_c/2}\langle M_\rho^{2(l+j)}\rangle C_{2l}=-\langle M_\rho^{2j}\rangle \label{cf}
\end{equation}
for all $j=1,2,\ldots,N_c/2$, in order to minimize the mean value $\langle\mathscr F\rangle$. {\color{black}Note that here the moments are the measured values from the experiment.} There is a simple relation between {\color{black}the moment for the radical distance $M^{2k}_\rho$ and the moment along a particular axis $M^{2k}$} \cite{PhysRevLett.108.233601},
\begin{equation}
\langle M_\rho^{2k}\rangle={\langle M^{2k}\rangle 2^{2k}\over\binom{2k}{k}}={2^{2k}\over\binom{2k}{k}}\int d(M^2)\bar F(M^2)M^{2k}.
\end{equation}

Within classical theory the ensemble is described by a joint non-negative probability distribution $\rho(M_y,M_z)$. Therefore, $\langle\mathscr F\rangle$ must remain non-negative since $\mathscr F\ge 0$ and hence
\begin{equation}
\langle\mathscr F\rangle=\int dM_ydM_z \rho(M_y,M_z)\mathscr F(\sqrt{M^2_z+M^2_y})\ge 0.
\end{equation}
Here the trial function $\mathscr F\ge 0$ acts as a local probe in $M_y-M_z$ phase space, projecting out a region, and testing the positivity of the joint probability distribution $\rho(M_y, M_z)$ in that region. Given a distribution function and its moments $\langle M_\rho^{2j}\rangle$, $\mathscr F$ is defined via Eq. (\ref{cf}), so that it is maximally sensitive to a potentially negative region $\rho(M_y,M_z)<0$.

Therefore, we calculate $\langle\mathscr F\rangle$ for the magnetic moment distribution of interest reported in Ref. \cite{NaMC} and plot the result versus the cutoff order $N_c$ in Fig.~3. For $N_c\ge 10$, we find $\langle\mathscr F\rangle<0$ which is impossible for a classical system where a positive joint probability distribution $\rho(M_y,M_z)\ge 0$ can be defined. When $N_c$ is increasing, $\langle \mathscr F\rangle$ monotonically approaches $-0.024$, and the standard deviation approaches $0.01$. To calculate the latter, we randomly select 150 times half of the data for calculating $\langle \mathscr F \rangle$. For comparison, we also plot $\langle\mathscr F\rangle$ for the (classically allowed) reference state with all atomic magnetic moments aligned, where we find always $\langle \mathscr F\rangle\ge 0$ as expected.
\begin{figure}
  \includegraphics[width=3in]{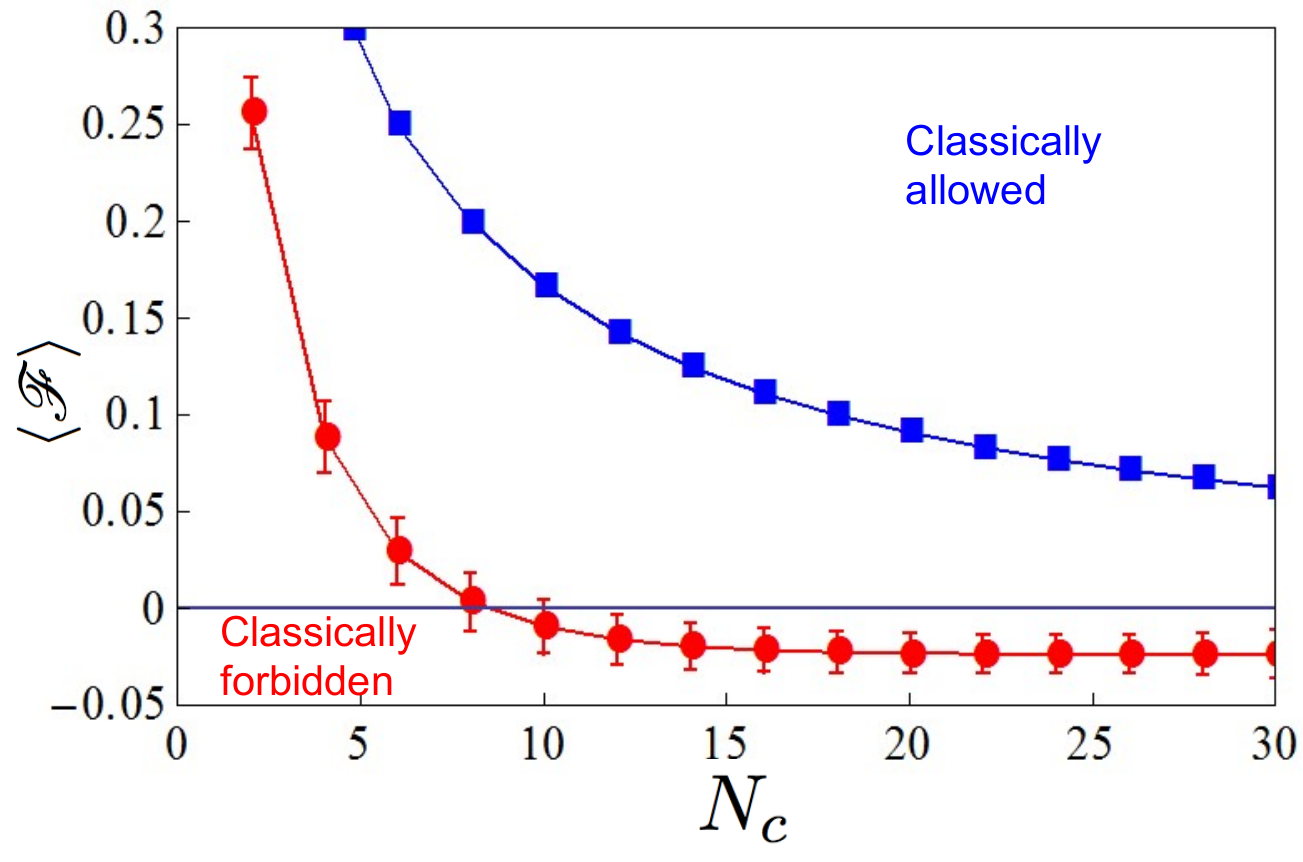}\\
  \caption{The mean value $\langle\mathscr F\rangle$ of the trial function versus the cutoff order $N_c$. The blue squares correspond to the reference state with all magnetic moments aligned, which has $\langle \mathscr F\rangle >0$, and does not violate classical physics. The analytic asymptotic limit for the reference state (blue squares) is $2/(N_c+2)$. The solid line here only joins the points in the plot. The red circles correspond to the state of interest. When $N_c$ is larger than 10, $\langle \mathscr F\rangle$ becomes negative, which is forbidden by classical physics. $\langle \mathscr F\rangle$ is monotonically decreasing and approaching -0.024 when $N_c$ is increasing. Here all error bars represent one standard deviation.}
\end{figure}

Compared to the previous analysis of the experiments \cite{NaMC} first reporting a negative Wigner function of an atomic ensemble, here we do not require any knowledge of the total spin, nor  do we involve the quantum formalism to define the quantum state \cite{reichel40atoms} and the Wigner function \cite{PhysRevA.49.4101}. Following Ref. \cite{PhysRevLett.108.233601}, we only use the marginal magnetic-moment distributions, which are insufficient to reconstruct the full quantum state, to demonstrate {\color{black}that} the observed ensemble magnetization cannot be explained with classical physics.

There is an interesting distinction between the violation found here for a large atomic system, and that observed in quantum optics for a single-photon Fock state \cite{PhysRevLett.108.233601,PhysRevLett.87.050402,RevModPhys.81.299}. In the case of quantum optics, the violation of the classical description is associated with a small structure in phase space of area $\Delta x\Delta p/\hbar\sim 1$, which is the level at which, according to the usual argument, quantum mechanics must be applied, {\color{black}as} the number of allowed states in this area is on the order of one. However, in the many-particle system we observe a much larger structure of area {\color{black}$\Delta M_y\Delta M_z/\mu^2_B\sim 10^3$} in phase space. Nevertheless {\color{black}this mesoscopic system} defies a classical description. It shows that the breakdown of the classical theory can be observed far above $\hbar$, the characteristic scale of quantum mechanics.

In conclusion, by analyzing the marginal magnetic moment distribution of an atomic ensemble with a total mass as large as $2.6\times 10^5$ amu, we verify the non-classical character of  the atomic magnetization distribution without using quantum mechanical assumptions about the atomic spin or magnetic moment, and with limited information that is insufficient to reconstruct the full quantum state. Remarkably, the detection of a single photon that has interacted with the atomic ensemble is sufficient to create a magnetization distribution that violates the laws of classical physics. This violation is ultimately a consequence of the fact that the magnetic moment cannot be simultaneously sharply defined along different directions. This, in turn, can affect the outcomes of mesoscopic or even macroscopic measurements.

This work was supported by NSF, DARPA, NASA, MURI grants through AFOSR and ARO and the European Union Seventh Framework Programme through the ERC Grant QIOS (Grant No. 306576).

\bibliographystyle{apsrev4-1}
\bibliography{violation}

\begin{thebibliography}{46}%
\makeatletter
\providecommand \@ifxundefined [1]{%
 \@ifx{#1\undefined}
}%
\providecommand \@ifnum [1]{%
 \ifnum #1\expandafter \@firstoftwo
 \else \expandafter \@secondoftwo
 \fi
}%
\providecommand \@ifx [1]{%
 \ifx #1\expandafter \@firstoftwo
 \else \expandafter \@secondoftwo
 \fi
}%
\providecommand \natexlab [1]{#1}%
\providecommand \enquote  [1]{``#1''}%
\providecommand \bibnamefont  [1]{#1}%
\providecommand \bibfnamefont [1]{#1}%
\providecommand \citenamefont [1]{#1}%
\providecommand \href@noop [0]{\@secondoftwo}%
\providecommand \href [0]{\begingroup \@sanitize@url \@href}%
\providecommand \@href[1]{\@@startlink{#1}\@@href}%
\providecommand \@@href[1]{\endgroup#1\@@endlink}%
\providecommand \@sanitize@url [0]{\catcode `\\12\catcode `\$12\catcode
  `\&12\catcode `\#12\catcode `\^12\catcode `\_12\catcode `\%12\relax}%
\providecommand \@@startlink[1]{}%
\providecommand \@@endlink[0]{}%
\providecommand \url  [0]{\begingroup\@sanitize@url \@url }%
\providecommand \@url [1]{\endgroup\@href {#1}{\urlprefix }}%
\providecommand \urlprefix  [0]{URL }%
\providecommand \Eprint [0]{\href }%
\providecommand \doibase [0]{http://dx.doi.org/}%
\providecommand \selectlanguage [0]{\@gobble}%
\providecommand \bibinfo  [0]{\@secondoftwo}%
\providecommand \bibfield  [0]{\@secondoftwo}%
\providecommand \translation [1]{[#1]}%
\providecommand \BibitemOpen [0]{}%
\providecommand \bibitemStop [0]{}%
\providecommand \bibitemNoStop [0]{.\EOS\space}%
\providecommand \EOS [0]{\spacefactor3000\relax}%
\providecommand \BibitemShut  [1]{\csname bibitem#1\endcsname}%
\let\auto@bib@innerbib\@empty
\bibitem [{\citenamefont {Bell}(1964)}]{bell}%
  \BibitemOpen
  \bibfield  {author} {\bibinfo {author} {\bibfnamefont {J.~S.}\ \bibnamefont
  {Bell}},\ }\href@noop {} {\bibfield  {journal} {\bibinfo  {journal}
  {Physics}\ }\textbf {\bibinfo {volume} {1 (3)}},\ \bibinfo {pages} {195}
  (\bibinfo {year} {1964})}\BibitemShut {NoStop}%
\bibitem [{\citenamefont {Clauser}\ \emph {et~al.}(1969)\citenamefont
  {Clauser}, \citenamefont {Horne}, \citenamefont {Shimony},\ and\
  \citenamefont {Holt}}]{CHSH}%
  \BibitemOpen
  \bibfield  {author} {\bibinfo {author} {\bibfnamefont {J.~F.}\ \bibnamefont
  {Clauser}}, \bibinfo {author} {\bibfnamefont {M.~A.}\ \bibnamefont {Horne}},
  \bibinfo {author} {\bibfnamefont {A.}~\bibnamefont {Shimony}}, \ and\
  \bibinfo {author} {\bibfnamefont {R.~A.}\ \bibnamefont {Holt}},\ }\href
  {\doibase 10.1103/PhysRevLett.23.880} {\bibfield  {journal} {\bibinfo
  {journal} {Phys. Rev. Lett.}\ }\textbf {\bibinfo {volume} {23}},\ \bibinfo
  {pages} {880} (\bibinfo {year} {1969})}\BibitemShut {NoStop}%
\bibitem [{\citenamefont {Aspect}\ \emph {et~al.}(1981)\citenamefont {Aspect},
  \citenamefont {Grangier},\ and\ \citenamefont {Roger}}]{aspect1}%
  \BibitemOpen
  \bibfield  {author} {\bibinfo {author} {\bibfnamefont {A.}~\bibnamefont
  {Aspect}}, \bibinfo {author} {\bibfnamefont {P.}~\bibnamefont {Grangier}}, \
  and\ \bibinfo {author} {\bibfnamefont {G.}~\bibnamefont {Roger}},\ }\href
  {\doibase 10.1103/PhysRevLett.47.460} {\bibfield  {journal} {\bibinfo
  {journal} {Phys. Rev. Lett.}\ }\textbf {\bibinfo {volume} {47}},\ \bibinfo
  {pages} {460} (\bibinfo {year} {1981})}\BibitemShut {NoStop}%
\bibitem [{\citenamefont {Aspect}\ \emph {et~al.}(1982)\citenamefont {Aspect},
  \citenamefont {Grangier},\ and\ \citenamefont {Roger}}]{aspect2}%
  \BibitemOpen
  \bibfield  {author} {\bibinfo {author} {\bibfnamefont {A.}~\bibnamefont
  {Aspect}}, \bibinfo {author} {\bibfnamefont {P.}~\bibnamefont {Grangier}}, \
  and\ \bibinfo {author} {\bibfnamefont {G.}~\bibnamefont {Roger}},\ }\href
  {\doibase 10.1103/PhysRevLett.49.91} {\bibfield  {journal} {\bibinfo
  {journal} {Phys. Rev. Lett.}\ }\textbf {\bibinfo {volume} {49}},\ \bibinfo
  {pages} {91} (\bibinfo {year} {1982})}\BibitemShut {NoStop}%
\bibitem [{\citenamefont {Giustina}\ \emph {et~al.}(2015)\citenamefont
  {Giustina}, \citenamefont {Versteegh}, \citenamefont {Wengerowsky},
  \citenamefont {Handsteiner}, \citenamefont {Hochrainer}, \citenamefont
  {Phelan}, \citenamefont {Steinlechner}, \citenamefont {Kofler}, \citenamefont
  {Larsson}, \citenamefont {Abell\'an}, \citenamefont {Amaya}, \citenamefont
  {Pruneri}, \citenamefont {Mitchell}, \citenamefont {Beyer}, \citenamefont
  {Gerrits}, \citenamefont {Lita}, \citenamefont {Shalm}, \citenamefont {Nam},
  \citenamefont {Scheidl}, \citenamefont {Ursin}, \citenamefont {Wittmann},\
  and\ \citenamefont {Zeilinger}}]{loop1}%
  \BibitemOpen
  \bibfield  {author} {\bibinfo {author} {\bibfnamefont {M.}~\bibnamefont
  {Giustina}}, \bibinfo {author} {\bibfnamefont {M.~A.~M.}\ \bibnamefont
  {Versteegh}}, \bibinfo {author} {\bibfnamefont {S.}~\bibnamefont
  {Wengerowsky}}, \bibinfo {author} {\bibfnamefont {J.}~\bibnamefont
  {Handsteiner}}, \bibinfo {author} {\bibfnamefont {A.}~\bibnamefont
  {Hochrainer}}, \bibinfo {author} {\bibfnamefont {K.}~\bibnamefont {Phelan}},
  \bibinfo {author} {\bibfnamefont {F.}~\bibnamefont {Steinlechner}}, \bibinfo
  {author} {\bibfnamefont {J.}~\bibnamefont {Kofler}}, \bibinfo {author}
  {\bibfnamefont {J.-A.}\ \bibnamefont {Larsson}}, \bibinfo {author}
  {\bibfnamefont {C.}~\bibnamefont {Abell\'an}}, \bibinfo {author}
  {\bibfnamefont {W.}~\bibnamefont {Amaya}}, \bibinfo {author} {\bibfnamefont
  {V.}~\bibnamefont {Pruneri}}, \bibinfo {author} {\bibfnamefont {M.~W.}\
  \bibnamefont {Mitchell}}, \bibinfo {author} {\bibfnamefont {J.}~\bibnamefont
  {Beyer}}, \bibinfo {author} {\bibfnamefont {T.}~\bibnamefont {Gerrits}},
  \bibinfo {author} {\bibfnamefont {A.~E.}\ \bibnamefont {Lita}}, \bibinfo
  {author} {\bibfnamefont {L.~K.}\ \bibnamefont {Shalm}}, \bibinfo {author}
  {\bibfnamefont {S.~W.}\ \bibnamefont {Nam}}, \bibinfo {author} {\bibfnamefont
  {T.}~\bibnamefont {Scheidl}}, \bibinfo {author} {\bibfnamefont
  {R.}~\bibnamefont {Ursin}}, \bibinfo {author} {\bibfnamefont
  {B.}~\bibnamefont {Wittmann}}, \ and\ \bibinfo {author} {\bibfnamefont
  {A.}~\bibnamefont {Zeilinger}},\ }\href {\doibase
  10.1103/PhysRevLett.115.250401} {\bibfield  {journal} {\bibinfo  {journal}
  {Phys. Rev. Lett.}\ }\textbf {\bibinfo {volume} {115}},\ \bibinfo {pages}
  {250401} (\bibinfo {year} {2015})}\BibitemShut {NoStop}%
\bibitem [{\citenamefont {Shalm}\ \emph {et~al.}(2015)\citenamefont {Shalm},
  \citenamefont {Meyer-Scott}, \citenamefont {Christensen}, \citenamefont
  {Bierhorst}, \citenamefont {Wayne}, \citenamefont {Stevens}, \citenamefont
  {Gerrits}, \citenamefont {Glancy}, \citenamefont {Hamel}, \citenamefont
  {Allman}, \citenamefont {Coakley}, \citenamefont {Dyer}, \citenamefont
  {Hodge}, \citenamefont {Lita}, \citenamefont {Verma}, \citenamefont
  {Lambrocco}, \citenamefont {Tortorici}, \citenamefont {Migdall},
  \citenamefont {Zhang}, \citenamefont {Kumor}, \citenamefont {Farr},
  \citenamefont {Marsili}, \citenamefont {Shaw}, \citenamefont {Stern},
  \citenamefont {Abell\'an}, \citenamefont {Amaya}, \citenamefont {Pruneri},
  \citenamefont {Jennewein}, \citenamefont {Mitchell}, \citenamefont {Kwiat},
  \citenamefont {Bienfang}, \citenamefont {Mirin}, \citenamefont {Knill},\ and\
  \citenamefont {Nam}}]{loop2}%
  \BibitemOpen
  \bibfield  {author} {\bibinfo {author} {\bibfnamefont {L.~K.}\ \bibnamefont
  {Shalm}}, \bibinfo {author} {\bibfnamefont {E.}~\bibnamefont {Meyer-Scott}},
  \bibinfo {author} {\bibfnamefont {B.~G.}\ \bibnamefont {Christensen}},
  \bibinfo {author} {\bibfnamefont {P.}~\bibnamefont {Bierhorst}}, \bibinfo
  {author} {\bibfnamefont {M.~A.}\ \bibnamefont {Wayne}}, \bibinfo {author}
  {\bibfnamefont {M.~J.}\ \bibnamefont {Stevens}}, \bibinfo {author}
  {\bibfnamefont {T.}~\bibnamefont {Gerrits}}, \bibinfo {author} {\bibfnamefont
  {S.}~\bibnamefont {Glancy}}, \bibinfo {author} {\bibfnamefont {D.~R.}\
  \bibnamefont {Hamel}}, \bibinfo {author} {\bibfnamefont {M.~S.}\ \bibnamefont
  {Allman}}, \bibinfo {author} {\bibfnamefont {K.~J.}\ \bibnamefont {Coakley}},
  \bibinfo {author} {\bibfnamefont {S.~D.}\ \bibnamefont {Dyer}}, \bibinfo
  {author} {\bibfnamefont {C.}~\bibnamefont {Hodge}}, \bibinfo {author}
  {\bibfnamefont {A.~E.}\ \bibnamefont {Lita}}, \bibinfo {author}
  {\bibfnamefont {V.~B.}\ \bibnamefont {Verma}}, \bibinfo {author}
  {\bibfnamefont {C.}~\bibnamefont {Lambrocco}}, \bibinfo {author}
  {\bibfnamefont {E.}~\bibnamefont {Tortorici}}, \bibinfo {author}
  {\bibfnamefont {A.~L.}\ \bibnamefont {Migdall}}, \bibinfo {author}
  {\bibfnamefont {Y.}~\bibnamefont {Zhang}}, \bibinfo {author} {\bibfnamefont
  {D.~R.}\ \bibnamefont {Kumor}}, \bibinfo {author} {\bibfnamefont {W.~H.}\
  \bibnamefont {Farr}}, \bibinfo {author} {\bibfnamefont {F.}~\bibnamefont
  {Marsili}}, \bibinfo {author} {\bibfnamefont {M.~D.}\ \bibnamefont {Shaw}},
  \bibinfo {author} {\bibfnamefont {J.~A.}\ \bibnamefont {Stern}}, \bibinfo
  {author} {\bibfnamefont {C.}~\bibnamefont {Abell\'an}}, \bibinfo {author}
  {\bibfnamefont {W.}~\bibnamefont {Amaya}}, \bibinfo {author} {\bibfnamefont
  {V.}~\bibnamefont {Pruneri}}, \bibinfo {author} {\bibfnamefont
  {T.}~\bibnamefont {Jennewein}}, \bibinfo {author} {\bibfnamefont {M.~W.}\
  \bibnamefont {Mitchell}}, \bibinfo {author} {\bibfnamefont {P.~G.}\
  \bibnamefont {Kwiat}}, \bibinfo {author} {\bibfnamefont {J.~C.}\ \bibnamefont
  {Bienfang}}, \bibinfo {author} {\bibfnamefont {R.~P.}\ \bibnamefont {Mirin}},
  \bibinfo {author} {\bibfnamefont {E.}~\bibnamefont {Knill}}, \ and\ \bibinfo
  {author} {\bibfnamefont {S.~W.}\ \bibnamefont {Nam}},\ }\href {\doibase
  10.1103/PhysRevLett.115.250402} {\bibfield  {journal} {\bibinfo  {journal}
  {Phys. Rev. Lett.}\ }\textbf {\bibinfo {volume} {115}},\ \bibinfo {pages}
  {250402} (\bibinfo {year} {2015})}\BibitemShut {NoStop}%
\bibitem [{\citenamefont {Schmied}\ \emph {et~al.}(2016)\citenamefont
  {Schmied}, \citenamefont {Bancal}, \citenamefont {Allard}, \citenamefont
  {Fadel}, \citenamefont {Scarani}, \citenamefont {Treutlein},\ and\
  \citenamefont {Sangouard}}]{Schmied441}%
  \BibitemOpen
  \bibfield  {author} {\bibinfo {author} {\bibfnamefont {R.}~\bibnamefont
  {Schmied}}, \bibinfo {author} {\bibfnamefont {J.-D.}\ \bibnamefont {Bancal}},
  \bibinfo {author} {\bibfnamefont {B.}~\bibnamefont {Allard}}, \bibinfo
  {author} {\bibfnamefont {M.}~\bibnamefont {Fadel}}, \bibinfo {author}
  {\bibfnamefont {V.}~\bibnamefont {Scarani}}, \bibinfo {author} {\bibfnamefont
  {P.}~\bibnamefont {Treutlein}}, \ and\ \bibinfo {author} {\bibfnamefont
  {N.}~\bibnamefont {Sangouard}},\ }\href {\doibase 10.1126/science.aad8665}
  {\bibfield  {journal} {\bibinfo  {journal} {Science}\ }\textbf {\bibinfo
  {volume} {352}},\ \bibinfo {pages} {441} (\bibinfo {year}
  {2016})}\BibitemShut {NoStop}%
\bibitem [{\citenamefont {Greenberger}\ \emph {et~al.}(1990)\citenamefont
  {Greenberger}, \citenamefont {Horne}, \citenamefont {Shimony},\ and\
  \citenamefont {Zeilinger}}]{GHZ}%
  \BibitemOpen
  \bibfield  {author} {\bibinfo {author} {\bibfnamefont {D.~M.}\ \bibnamefont
  {Greenberger}}, \bibinfo {author} {\bibfnamefont {M.~A.}\ \bibnamefont
  {Horne}}, \bibinfo {author} {\bibfnamefont {A.}~\bibnamefont {Shimony}}, \
  and\ \bibinfo {author} {\bibfnamefont {A.}~\bibnamefont {Zeilinger}},\ }\href
  {\doibase http://dx.doi.org/10.1119/1.16243} {\bibfield  {journal} {\bibinfo
  {journal} {American Journal of Physics}\ }\textbf {\bibinfo {volume} {58}},\
  \bibinfo {pages} {1131} (\bibinfo {year} {1990})}\BibitemShut {NoStop}%
\bibitem [{\citenamefont {Leibfried}\ \emph {et~al.}(2004)\citenamefont
  {Leibfried}, \citenamefont {Barrett}, \citenamefont {Schaetz}, \citenamefont
  {Britton}, \citenamefont {Chiaverini}, \citenamefont {Itano}, \citenamefont
  {Jost}, \citenamefont {Langer},\ and\ \citenamefont
  {Wineland}}]{Leibfried04062004}%
  \BibitemOpen
  \bibfield  {author} {\bibinfo {author} {\bibfnamefont {D.}~\bibnamefont
  {Leibfried}}, \bibinfo {author} {\bibfnamefont {M.~D.}\ \bibnamefont
  {Barrett}}, \bibinfo {author} {\bibfnamefont {T.}~\bibnamefont {Schaetz}},
  \bibinfo {author} {\bibfnamefont {J.}~\bibnamefont {Britton}}, \bibinfo
  {author} {\bibfnamefont {J.}~\bibnamefont {Chiaverini}}, \bibinfo {author}
  {\bibfnamefont {W.~M.}\ \bibnamefont {Itano}}, \bibinfo {author}
  {\bibfnamefont {J.~D.}\ \bibnamefont {Jost}}, \bibinfo {author}
  {\bibfnamefont {C.}~\bibnamefont {Langer}}, \ and\ \bibinfo {author}
  {\bibfnamefont {D.~J.}\ \bibnamefont {Wineland}},\ }\href {\doibase
  10.1126/science.1097576} {\bibfield  {journal} {\bibinfo  {journal}
  {Science}\ }\textbf {\bibinfo {volume} {304}},\ \bibinfo {pages} {1476}
  (\bibinfo {year} {2004})}\BibitemShut {NoStop}%
\bibitem [{\citenamefont {Roos}\ \emph {et~al.}(2004)\citenamefont {Roos},
  \citenamefont {Riebe}, \citenamefont {H\"affner}, \citenamefont {H\"ansel},
  \citenamefont {Benhelm}, \citenamefont {Lancaster}, \citenamefont {Becher},
  \citenamefont {Schmidt-Kaler},\ and\ \citenamefont {Blatt}}]{Roos04062004}%
  \BibitemOpen
  \bibfield  {author} {\bibinfo {author} {\bibfnamefont {C.~F.}\ \bibnamefont
  {Roos}}, \bibinfo {author} {\bibfnamefont {M.}~\bibnamefont {Riebe}},
  \bibinfo {author} {\bibfnamefont {H.}~\bibnamefont {H\"affner}}, \bibinfo
  {author} {\bibfnamefont {W.}~\bibnamefont {H\"ansel}}, \bibinfo {author}
  {\bibfnamefont {J.}~\bibnamefont {Benhelm}}, \bibinfo {author} {\bibfnamefont
  {G.~P.~T.}\ \bibnamefont {Lancaster}}, \bibinfo {author} {\bibfnamefont
  {C.}~\bibnamefont {Becher}}, \bibinfo {author} {\bibfnamefont
  {F.}~\bibnamefont {Schmidt-Kaler}}, \ and\ \bibinfo {author} {\bibfnamefont
  {R.}~\bibnamefont {Blatt}},\ }\href {\doibase 10.1126/science.1097522}
  {\bibfield  {journal} {\bibinfo  {journal} {Science}\ }\textbf {\bibinfo
  {volume} {304}},\ \bibinfo {pages} {1478} (\bibinfo {year}
  {2004})}\BibitemShut {NoStop}%
\bibitem [{\citenamefont {Monz}\ \emph {et~al.}(2011)\citenamefont {Monz},
  \citenamefont {Schindler}, \citenamefont {Barreiro}, \citenamefont {Chwalla},
  \citenamefont {Nigg}, \citenamefont {Coish}, \citenamefont {Harlander},
  \citenamefont {H\"ansel}, \citenamefont {Hennrich},\ and\ \citenamefont
  {Blatt}}]{PhysRevLett.106.130506}%
  \BibitemOpen
  \bibfield  {author} {\bibinfo {author} {\bibfnamefont {T.}~\bibnamefont
  {Monz}}, \bibinfo {author} {\bibfnamefont {P.}~\bibnamefont {Schindler}},
  \bibinfo {author} {\bibfnamefont {J.~T.}\ \bibnamefont {Barreiro}}, \bibinfo
  {author} {\bibfnamefont {M.}~\bibnamefont {Chwalla}}, \bibinfo {author}
  {\bibfnamefont {D.}~\bibnamefont {Nigg}}, \bibinfo {author} {\bibfnamefont
  {W.~A.}\ \bibnamefont {Coish}}, \bibinfo {author} {\bibfnamefont
  {M.}~\bibnamefont {Harlander}}, \bibinfo {author} {\bibfnamefont
  {W.}~\bibnamefont {H\"ansel}}, \bibinfo {author} {\bibfnamefont
  {M.}~\bibnamefont {Hennrich}}, \ and\ \bibinfo {author} {\bibfnamefont
  {R.}~\bibnamefont {Blatt}},\ }\href {\doibase 10.1103/PhysRevLett.106.130506}
  {\bibfield  {journal} {\bibinfo  {journal} {Phys. Rev. Lett.}\ }\textbf
  {\bibinfo {volume} {106}},\ \bibinfo {pages} {130506} (\bibinfo {year}
  {2011})}\BibitemShut {NoStop}%
\bibitem [{\citenamefont {Gerlich}\ \emph {et~al.}(2011)\citenamefont
  {Gerlich}, \citenamefont {Eibenberger}, \citenamefont {Tomandl},
  \citenamefont {Nimmrichter}, \citenamefont {Hornberger}, \citenamefont
  {Fagan}, \citenamefont {Tuxen}, \citenamefont {Mayor},\ and\ \citenamefont
  {Arndt}}]{Size4}%
  \BibitemOpen
  \bibfield  {author} {\bibinfo {author} {\bibfnamefont {S.}~\bibnamefont
  {Gerlich}}, \bibinfo {author} {\bibfnamefont {S.}~\bibnamefont
  {Eibenberger}}, \bibinfo {author} {\bibfnamefont {M.}~\bibnamefont
  {Tomandl}}, \bibinfo {author} {\bibfnamefont {S.}~\bibnamefont
  {Nimmrichter}}, \bibinfo {author} {\bibfnamefont {K.}~\bibnamefont
  {Hornberger}}, \bibinfo {author} {\bibfnamefont {P.~J.}\ \bibnamefont
  {Fagan}}, \bibinfo {author} {\bibfnamefont {J.}~\bibnamefont {Tuxen}},
  \bibinfo {author} {\bibfnamefont {M.}~\bibnamefont {Mayor}}, \ and\ \bibinfo
  {author} {\bibfnamefont {M.}~\bibnamefont {Arndt}},\ }\href@noop {}
  {\bibfield  {journal} {\bibinfo  {journal} {Nature Communications}\ }\textbf
  {\bibinfo {volume} {2}},\ \bibinfo {pages} {263} (\bibinfo {year}
  {2011})}\BibitemShut {NoStop}%
\bibitem [{\citenamefont {Arndt}\ and\ \citenamefont
  {Hornberger}(2014)}]{Size1}%
  \BibitemOpen
  \bibfield  {author} {\bibinfo {author} {\bibfnamefont {M.}~\bibnamefont
  {Arndt}}\ and\ \bibinfo {author} {\bibfnamefont {K.}~\bibnamefont
  {Hornberger}},\ }\href@noop {} {\bibfield  {journal} {\bibinfo  {journal}
  {Nature Physics}\ }\textbf {\bibinfo {volume} {10}},\ \bibinfo {pages} {271}
  (\bibinfo {year} {2014})}\BibitemShut {NoStop}%
\bibitem [{\citenamefont {Hornberger}\ \emph {et~al.}(2012)\citenamefont
  {Hornberger}, \citenamefont {Gerlich}, \citenamefont {Haslinger},
  \citenamefont {Nimmrichter},\ and\ \citenamefont {Arndt}}]{Size2}%
  \BibitemOpen
  \bibfield  {author} {\bibinfo {author} {\bibfnamefont {K.}~\bibnamefont
  {Hornberger}}, \bibinfo {author} {\bibfnamefont {S.}~\bibnamefont {Gerlich}},
  \bibinfo {author} {\bibfnamefont {P.}~\bibnamefont {Haslinger}}, \bibinfo
  {author} {\bibfnamefont {S.}~\bibnamefont {Nimmrichter}}, \ and\ \bibinfo
  {author} {\bibfnamefont {M.}~\bibnamefont {Arndt}},\ }\href {\doibase
  10.1103/RevModPhys.84.157} {\bibfield  {journal} {\bibinfo  {journal} {Rev.
  Mod. Phys.}\ }\textbf {\bibinfo {volume} {84}},\ \bibinfo {pages} {157}
  (\bibinfo {year} {2012})}\BibitemShut {NoStop}%
\bibitem [{\citenamefont {D\"orre}\ \emph {et~al.}(2014)\citenamefont
  {D\"orre}, \citenamefont {Rodewald}, \citenamefont {Geyer}, \citenamefont
  {von Issendorff}, \citenamefont {Haslinger},\ and\ \citenamefont
  {Arndt}}]{Size3}%
  \BibitemOpen
  \bibfield  {author} {\bibinfo {author} {\bibfnamefont {N.}~\bibnamefont
  {D\"orre}}, \bibinfo {author} {\bibfnamefont {J.}~\bibnamefont {Rodewald}},
  \bibinfo {author} {\bibfnamefont {P.}~\bibnamefont {Geyer}}, \bibinfo
  {author} {\bibfnamefont {B.}~\bibnamefont {von Issendorff}}, \bibinfo
  {author} {\bibfnamefont {P.}~\bibnamefont {Haslinger}}, \ and\ \bibinfo
  {author} {\bibfnamefont {M.}~\bibnamefont {Arndt}},\ }\href {\doibase
  10.1103/PhysRevLett.113.233001} {\bibfield  {journal} {\bibinfo  {journal}
  {Phys. Rev. Lett.}\ }\textbf {\bibinfo {volume} {113}},\ \bibinfo {pages}
  {233001} (\bibinfo {year} {2014})}\BibitemShut {NoStop}%
\bibitem [{\citenamefont {Wigner}(1932)}]{WF}%
  \BibitemOpen
  \bibfield  {author} {\bibinfo {author} {\bibfnamefont {E.}~\bibnamefont
  {Wigner}},\ }\href {\doibase 10.1103/PhysRev.40.749} {\bibfield  {journal}
  {\bibinfo  {journal} {Phys. Rev.}\ }\textbf {\bibinfo {volume} {40}},\
  \bibinfo {pages} {749} (\bibinfo {year} {1932})}\BibitemShut {NoStop}%
\bibitem [{\citenamefont {Terhal}(2000)}]{Terhal2000319}%
  \BibitemOpen
  \bibfield  {author} {\bibinfo {author} {\bibfnamefont {B.~M.}\ \bibnamefont
  {Terhal}},\ }\href {\doibase http://dx.doi.org/10.1016/S0375-9601(00)00401-1}
  {\bibfield  {journal} {\bibinfo  {journal} {Physics Letters A}\ }\textbf
  {\bibinfo {volume} {271}},\ \bibinfo {pages} {319 } (\bibinfo {year}
  {2000})}\BibitemShut {NoStop}%
\bibitem [{\citenamefont {Bourennane}\ \emph {et~al.}(2004)\citenamefont
  {Bourennane}, \citenamefont {Eibl}, \citenamefont {Kurtsiefer}, \citenamefont
  {Gaertner}, \citenamefont {Weinfurter}, \citenamefont {G\"uhne},
  \citenamefont {Hyllus}, \citenamefont {Bru\ss{}}, \citenamefont
  {Lewenstein},\ and\ \citenamefont {Sanpera}}]{PhysRevLett.92.087902}%
  \BibitemOpen
  \bibfield  {author} {\bibinfo {author} {\bibfnamefont {M.}~\bibnamefont
  {Bourennane}}, \bibinfo {author} {\bibfnamefont {M.}~\bibnamefont {Eibl}},
  \bibinfo {author} {\bibfnamefont {C.}~\bibnamefont {Kurtsiefer}}, \bibinfo
  {author} {\bibfnamefont {S.}~\bibnamefont {Gaertner}}, \bibinfo {author}
  {\bibfnamefont {H.}~\bibnamefont {Weinfurter}}, \bibinfo {author}
  {\bibfnamefont {O.}~\bibnamefont {G\"uhne}}, \bibinfo {author} {\bibfnamefont
  {P.}~\bibnamefont {Hyllus}}, \bibinfo {author} {\bibfnamefont
  {D.}~\bibnamefont {Bru\ss{}}}, \bibinfo {author} {\bibfnamefont
  {M.}~\bibnamefont {Lewenstein}}, \ and\ \bibinfo {author} {\bibfnamefont
  {A.}~\bibnamefont {Sanpera}},\ }\href {\doibase
  10.1103/PhysRevLett.92.087902} {\bibfield  {journal} {\bibinfo  {journal}
  {Phys. Rev. Lett.}\ }\textbf {\bibinfo {volume} {92}},\ \bibinfo {pages}
  {087902} (\bibinfo {year} {2004})}\BibitemShut {NoStop}%
\bibitem [{\citenamefont {Simon}(2000)}]{PhysRevLett.84.2726}%
  \BibitemOpen
  \bibfield  {author} {\bibinfo {author} {\bibfnamefont {R.}~\bibnamefont
  {Simon}},\ }\href {\doibase 10.1103/PhysRevLett.84.2726} {\bibfield
  {journal} {\bibinfo  {journal} {Phys. Rev. Lett.}\ }\textbf {\bibinfo
  {volume} {84}},\ \bibinfo {pages} {2726} (\bibinfo {year}
  {2000})}\BibitemShut {NoStop}%
\bibitem [{\citenamefont {Duan}\ \emph {et~al.}(2000)\citenamefont {Duan},
  \citenamefont {Giedke}, \citenamefont {Cirac},\ and\ \citenamefont
  {Zoller}}]{PhysRevLett.84.2722}%
  \BibitemOpen
  \bibfield  {author} {\bibinfo {author} {\bibfnamefont {L.-M.}\ \bibnamefont
  {Duan}}, \bibinfo {author} {\bibfnamefont {G.}~\bibnamefont {Giedke}},
  \bibinfo {author} {\bibfnamefont {J.~I.}\ \bibnamefont {Cirac}}, \ and\
  \bibinfo {author} {\bibfnamefont {P.}~\bibnamefont {Zoller}},\ }\href
  {\doibase 10.1103/PhysRevLett.84.2722} {\bibfield  {journal} {\bibinfo
  {journal} {Phys. Rev. Lett.}\ }\textbf {\bibinfo {volume} {84}},\ \bibinfo
  {pages} {2722} (\bibinfo {year} {2000})}\BibitemShut {NoStop}%
\bibitem [{\citenamefont {Gr\o{}nbech-Jensen}\ \emph
  {et~al.}(2010)\citenamefont {Gr\o{}nbech-Jensen}, \citenamefont {Marchese},
  \citenamefont {Cirillo},\ and\ \citenamefont
  {Blackburn}}]{PhysRevLett.105.010501}%
  \BibitemOpen
  \bibfield  {author} {\bibinfo {author} {\bibfnamefont {N.}~\bibnamefont
  {Gr\o{}nbech-Jensen}}, \bibinfo {author} {\bibfnamefont {J.~E.}\ \bibnamefont
  {Marchese}}, \bibinfo {author} {\bibfnamefont {M.}~\bibnamefont {Cirillo}}, \
  and\ \bibinfo {author} {\bibfnamefont {J.~A.}\ \bibnamefont {Blackburn}},\
  }\href {\doibase 10.1103/PhysRevLett.105.010501} {\bibfield  {journal}
  {\bibinfo  {journal} {Phys. Rev. Lett.}\ }\textbf {\bibinfo {volume} {105}},\
  \bibinfo {pages} {010501} (\bibinfo {year} {2010})}\BibitemShut {NoStop}%
\bibitem [{\citenamefont {Kot}\ \emph {et~al.}(2012)\citenamefont {Kot},
  \citenamefont {Gr\o{}nbech-Jensen}, \citenamefont {Nielsen}, \citenamefont
  {Neergaard-Nielsen}, \citenamefont {Polzik},\ and\ \citenamefont
  {S\o{}rensen}}]{PhysRevLett.108.233601}%
  \BibitemOpen
  \bibfield  {author} {\bibinfo {author} {\bibfnamefont {E.}~\bibnamefont
  {Kot}}, \bibinfo {author} {\bibfnamefont {N.}~\bibnamefont
  {Gr\o{}nbech-Jensen}}, \bibinfo {author} {\bibfnamefont {B.~M.}\ \bibnamefont
  {Nielsen}}, \bibinfo {author} {\bibfnamefont {J.~S.}\ \bibnamefont
  {Neergaard-Nielsen}}, \bibinfo {author} {\bibfnamefont {E.~S.}\ \bibnamefont
  {Polzik}}, \ and\ \bibinfo {author} {\bibfnamefont {A.~S.}\ \bibnamefont
  {S\o{}rensen}},\ }\href {\doibase 10.1103/PhysRevLett.108.233601} {\bibfield
  {journal} {\bibinfo  {journal} {Phys. Rev. Lett.}\ }\textbf {\bibinfo
  {volume} {108}},\ \bibinfo {pages} {233601} (\bibinfo {year}
  {2012})}\BibitemShut {NoStop}%
\bibitem [{\citenamefont {Carmichael}\ and\ \citenamefont
  {Walls}(1976)}]{anti2}%
  \BibitemOpen
  \bibfield  {author} {\bibinfo {author} {\bibfnamefont {H.~J.}\ \bibnamefont
  {Carmichael}}\ and\ \bibinfo {author} {\bibfnamefont {D.~F.}\ \bibnamefont
  {Walls}},\ }\href {http://stacks.iop.org/0022-3700/9/i=8/a=007} {\bibfield
  {journal} {\bibinfo  {journal} {Journal of Physics B: Atomic and Molecular
  Physics}\ }\textbf {\bibinfo {volume} {9}},\ \bibinfo {pages} {1199}
  (\bibinfo {year} {1976})}\BibitemShut {NoStop}%
\bibitem [{\citenamefont {Kimble}\ and\ \citenamefont {Mandel}(1976)}]{anti3}%
  \BibitemOpen
  \bibfield  {author} {\bibinfo {author} {\bibfnamefont {H.~J.}\ \bibnamefont
  {Kimble}}\ and\ \bibinfo {author} {\bibfnamefont {L.}~\bibnamefont
  {Mandel}},\ }\href {\doibase 10.1103/PhysRevA.13.2123} {\bibfield  {journal}
  {\bibinfo  {journal} {Phys. Rev. A}\ }\textbf {\bibinfo {volume} {13}},\
  \bibinfo {pages} {2123} (\bibinfo {year} {1976})}\BibitemShut {NoStop}%
\bibitem [{\citenamefont {Klyshko}(1994)}]{Klyshko1}%
  \BibitemOpen
  \bibfield  {author} {\bibinfo {author} {\bibfnamefont {D.~N.}\ \bibnamefont
  {Klyshko}},\ }\href {\doibase 10.1070/PU1994v037n11ABEH000054} {\bibfield
  {journal} {\bibinfo  {journal} {Physics-Uspekhi}\ }\textbf {\bibinfo {volume}
  {37}},\ \bibinfo {pages} {1097} (\bibinfo {year} {1994})}\BibitemShut
  {NoStop}%
\bibitem [{\citenamefont {Klyshko}(1996)}]{Klyshko2}%
  \BibitemOpen
  \bibfield  {author} {\bibinfo {author} {\bibfnamefont {D.}~\bibnamefont
  {Klyshko}},\ }\href {\doibase http://dx.doi.org/10.1016/0375-9601(96)00091-6}
  {\bibfield  {journal} {\bibinfo  {journal} {Physics Letters A}\ }\textbf
  {\bibinfo {volume} {213}},\ \bibinfo {pages} {7 } (\bibinfo {year}
  {1996})}\BibitemShut {NoStop}%
\bibitem [{\citenamefont {Mandel}(1986)}]{Sta1}%
  \BibitemOpen
  \bibfield  {author} {\bibinfo {author} {\bibfnamefont {L.}~\bibnamefont
  {Mandel}},\ }\href {http://stacks.iop.org/1402-4896/1986/i=T12/a=005}
  {\bibfield  {journal} {\bibinfo  {journal} {Physica Scripta}\ }\textbf
  {\bibinfo {volume} {1986}},\ \bibinfo {pages} {34} (\bibinfo {year}
  {1986})}\BibitemShut {NoStop}%
\bibitem [{\citenamefont {Agarwal}(1993)}]{Sta2}%
  \BibitemOpen
  \bibfield  {author} {\bibinfo {author} {\bibfnamefont {G.}~\bibnamefont
  {Agarwal}},\ }\href {\doibase http://dx.doi.org/10.1016/0030-4018(93)90059-E}
  {\bibfield  {journal} {\bibinfo  {journal} {Optics Communications}\ }\textbf
  {\bibinfo {volume} {95}},\ \bibinfo {pages} {109 } (\bibinfo {year}
  {1993})}\BibitemShut {NoStop}%
\bibitem [{\citenamefont {Agarwal}\ and\ \citenamefont {Tara}(1992)}]{Sta3}%
  \BibitemOpen
  \bibfield  {author} {\bibinfo {author} {\bibfnamefont {G.~S.}\ \bibnamefont
  {Agarwal}}\ and\ \bibinfo {author} {\bibfnamefont {K.}~\bibnamefont {Tara}},\
  }\href {\doibase 10.1103/PhysRevA.46.485} {\bibfield  {journal} {\bibinfo
  {journal} {Phys. Rev. A}\ }\textbf {\bibinfo {volume} {46}},\ \bibinfo
  {pages} {485} (\bibinfo {year} {1992})}\BibitemShut {NoStop}%
\bibitem [{\citenamefont {Richter}\ and\ \citenamefont {Vogel}(2002)}]{Sta4}%
  \BibitemOpen
  \bibfield  {author} {\bibinfo {author} {\bibfnamefont {T.}~\bibnamefont
  {Richter}}\ and\ \bibinfo {author} {\bibfnamefont {W.}~\bibnamefont
  {Vogel}},\ }\href {\doibase 10.1103/PhysRevLett.89.283601} {\bibfield
  {journal} {\bibinfo  {journal} {Phys. Rev. Lett.}\ }\textbf {\bibinfo
  {volume} {89}},\ \bibinfo {pages} {283601} (\bibinfo {year}
  {2002})}\BibitemShut {NoStop}%
\bibitem [{\citenamefont {Kimble}\ \emph {et~al.}(1977)\citenamefont {Kimble},
  \citenamefont {Dagenais},\ and\ \citenamefont {Mandel}}]{anti1}%
  \BibitemOpen
  \bibfield  {author} {\bibinfo {author} {\bibfnamefont {H.~J.}\ \bibnamefont
  {Kimble}}, \bibinfo {author} {\bibfnamefont {M.}~\bibnamefont {Dagenais}}, \
  and\ \bibinfo {author} {\bibfnamefont {L.}~\bibnamefont {Mandel}},\ }\href
  {\doibase 10.1103/PhysRevLett.39.691} {\bibfield  {journal} {\bibinfo
  {journal} {Phys. Rev. Lett.}\ }\textbf {\bibinfo {volume} {39}},\ \bibinfo
  {pages} {691} (\bibinfo {year} {1977})}\BibitemShut {NoStop}%
\bibitem [{\citenamefont {Short}\ and\ \citenamefont {Mandel}(1983)}]{Sta5}%
  \BibitemOpen
  \bibfield  {author} {\bibinfo {author} {\bibfnamefont {R.}~\bibnamefont
  {Short}}\ and\ \bibinfo {author} {\bibfnamefont {L.}~\bibnamefont {Mandel}},\
  }\href {\doibase 10.1103/PhysRevLett.51.384} {\bibfield  {journal} {\bibinfo
  {journal} {Phys. Rev. Lett.}\ }\textbf {\bibinfo {volume} {51}},\ \bibinfo
  {pages} {384} (\bibinfo {year} {1983})}\BibitemShut {NoStop}%
\bibitem [{\citenamefont {Zeilinger}(1999)}]{RMPZ}%
  \BibitemOpen
  \bibfield  {author} {\bibinfo {author} {\bibfnamefont {A.}~\bibnamefont
  {Zeilinger}},\ }\href {\doibase 10.1103/RevModPhys.71.S288} {\bibfield
  {journal} {\bibinfo  {journal} {Rev. Mod. Phys.}\ }\textbf {\bibinfo {volume}
  {71}},\ \bibinfo {pages} {S288} (\bibinfo {year} {1999})}\BibitemShut
  {NoStop}%
\bibitem [{\citenamefont {Knee}\ \emph {et~al.}(2016)\citenamefont {Knee},
  \citenamefont {Kakuyanagi}, \citenamefont {Yeh}, \citenamefont {Matsuzaki},
  \citenamefont {Toida}, \citenamefont {Yamaguchi}, \citenamefont {Saito},
  \citenamefont {Leggett},\ and\ \citenamefont {Munro}}]{Knee}%
  \BibitemOpen
  \bibfield  {author} {\bibinfo {author} {\bibfnamefont {G.~C.}\ \bibnamefont
  {Knee}}, \bibinfo {author} {\bibfnamefont {K.}~\bibnamefont {Kakuyanagi}},
  \bibinfo {author} {\bibfnamefont {M.-C.}\ \bibnamefont {Yeh}}, \bibinfo
  {author} {\bibfnamefont {Y.}~\bibnamefont {Matsuzaki}}, \bibinfo {author}
  {\bibfnamefont {H.}~\bibnamefont {Toida}}, \bibinfo {author} {\bibfnamefont
  {H.}~\bibnamefont {Yamaguchi}}, \bibinfo {author} {\bibfnamefont
  {S.}~\bibnamefont {Saito}}, \bibinfo {author} {\bibfnamefont {A.~J.}\
  \bibnamefont {Leggett}}, \ and\ \bibinfo {author} {\bibfnamefont {W.~J.}\
  \bibnamefont {Munro}},\ }\href {http://arxiv.org/abs/1601.03728} {\bibfield
  {journal} {\bibinfo  {journal} {e-print arXiv:1601.03728}\ } (\bibinfo {year}
  {2016})}\BibitemShut {NoStop}%
\bibitem [{\citenamefont {Vogel}(2000)}]{PhysRevLett.84.1849}%
  \BibitemOpen
  \bibfield  {author} {\bibinfo {author} {\bibfnamefont {W.}~\bibnamefont
  {Vogel}},\ }\href {\doibase 10.1103/PhysRevLett.84.1849} {\bibfield
  {journal} {\bibinfo  {journal} {Phys. Rev. Lett.}\ }\textbf {\bibinfo
  {volume} {84}},\ \bibinfo {pages} {1849} (\bibinfo {year}
  {2000})}\BibitemShut {NoStop}%
\bibitem [{\citenamefont {Lvovsky}\ and\ \citenamefont
  {Shapiro}(2002)}]{PhysRevA.65.033830}%
  \BibitemOpen
  \bibfield  {author} {\bibinfo {author} {\bibfnamefont {A.~I.}\ \bibnamefont
  {Lvovsky}}\ and\ \bibinfo {author} {\bibfnamefont {J.~H.}\ \bibnamefont
  {Shapiro}},\ }\href {\doibase 10.1103/PhysRevA.65.033830} {\bibfield
  {journal} {\bibinfo  {journal} {Phys. Rev. A}\ }\textbf {\bibinfo {volume}
  {65}},\ \bibinfo {pages} {033830} (\bibinfo {year} {2002})}\BibitemShut
  {NoStop}%
\bibitem [{\citenamefont {Bednorz}\ and\ \citenamefont
  {Belzig}(2011)}]{PhysRevA.83.052113}%
  \BibitemOpen
  \bibfield  {author} {\bibinfo {author} {\bibfnamefont {A.}~\bibnamefont
  {Bednorz}}\ and\ \bibinfo {author} {\bibfnamefont {W.}~\bibnamefont
  {Belzig}},\ }\href {\doibase 10.1103/PhysRevA.83.052113} {\bibfield
  {journal} {\bibinfo  {journal} {Phys. Rev. A}\ }\textbf {\bibinfo {volume}
  {83}},\ \bibinfo {pages} {052113} (\bibinfo {year} {2011})}\BibitemShut
  {NoStop}%
\bibitem [{\citenamefont {Kiesel}\ \emph {et~al.}(2012)\citenamefont {Kiesel},
  \citenamefont {Vogel}, \citenamefont {Christensen}, \citenamefont {B\'eguin},
  \citenamefont {Appel},\ and\ \citenamefont {Polzik}}]{PhysRevA.86.042108}%
  \BibitemOpen
  \bibfield  {author} {\bibinfo {author} {\bibfnamefont {T.}~\bibnamefont
  {Kiesel}}, \bibinfo {author} {\bibfnamefont {W.}~\bibnamefont {Vogel}},
  \bibinfo {author} {\bibfnamefont {S.~L.}\ \bibnamefont {Christensen}},
  \bibinfo {author} {\bibfnamefont {J.-B.}\ \bibnamefont {B\'eguin}}, \bibinfo
  {author} {\bibfnamefont {J.}~\bibnamefont {Appel}}, \ and\ \bibinfo {author}
  {\bibfnamefont {E.~S.}\ \bibnamefont {Polzik}},\ }\href {\doibase
  10.1103/PhysRevA.86.042108} {\bibfield  {journal} {\bibinfo  {journal} {Phys.
  Rev. A}\ }\textbf {\bibinfo {volume} {86}},\ \bibinfo {pages} {042108}
  (\bibinfo {year} {2012})}\BibitemShut {NoStop}%
\bibitem [{\citenamefont {Fresta}\ \emph {et~al.}(2015)\citenamefont {Fresta},
  \citenamefont {Borregaard},\ and\ \citenamefont {S\o{}rensen}}]{generaltest}%
  \BibitemOpen
  \bibfield  {author} {\bibinfo {author} {\bibfnamefont {L.}~\bibnamefont
  {Fresta}}, \bibinfo {author} {\bibfnamefont {J.}~\bibnamefont {Borregaard}},
  \ and\ \bibinfo {author} {\bibfnamefont {A.~S.}\ \bibnamefont
  {S\o{}rensen}},\ }\href {\doibase 10.1103/PhysRevA.92.062111} {\bibfield
  {journal} {\bibinfo  {journal} {Phys. Rev. A}\ }\textbf {\bibinfo {volume}
  {92}},\ \bibinfo {pages} {062111} (\bibinfo {year} {2015})}\BibitemShut
  {NoStop}%
\bibitem [{\citenamefont {McConnell}\ \emph {et~al.}(2015)\citenamefont
  {McConnell}, \citenamefont {Zhang}, \citenamefont {Hu}, \citenamefont
  {\ifmmode~\acute{C}\else \'{C}\fi{}uk},\ and\ \citenamefont
  {Vuleti\ifmmode~\acute{c}\else \'{c}\fi{}}}]{NaMC}%
  \BibitemOpen
  \bibfield  {author} {\bibinfo {author} {\bibfnamefont {R.}~\bibnamefont
  {McConnell}}, \bibinfo {author} {\bibfnamefont {H.}~\bibnamefont {Zhang}},
  \bibinfo {author} {\bibfnamefont {J.}~\bibnamefont {Hu}}, \bibinfo {author}
  {\bibfnamefont {S.}~\bibnamefont {\ifmmode~\acute{C}\else \'{C}\fi{}uk}}, \
  and\ \bibinfo {author} {\bibfnamefont {V.}~\bibnamefont
  {Vuleti\ifmmode~\acute{c}\else \'{c}\fi{}}},\ }\href@noop {} {\bibfield
  {journal} {\bibinfo  {journal} {Nature}\ }\textbf {\bibinfo {volume} {519}},\
  \bibinfo {pages} {439} (\bibinfo {year} {2015})}\BibitemShut {NoStop}%
\bibitem [{Note1()}]{Note1}%
  \BibitemOpen
  \bibinfo {note} {Note that the PSD criterion requires no knowledge of the
  total spin, and therefore many different quantum states are potentially
  consistent with the measured phase space distribution.}\BibitemShut {Stop}%
\bibitem [{\citenamefont {McConnell}\ \emph {et~al.}(2013)\citenamefont
  {McConnell}, \citenamefont {Zhang}, \citenamefont {\ifmmode~\acute{C}\else
  \'{C}\fi{}uk}, \citenamefont {Hu}, \citenamefont {Schleier-Smith},\ and\
  \citenamefont {Vuleti\ifmmode~\acute{c}\else
  \'{c}\fi{}}}]{PhysRevA.88.063802}%
  \BibitemOpen
  \bibfield  {author} {\bibinfo {author} {\bibfnamefont {R.}~\bibnamefont
  {McConnell}}, \bibinfo {author} {\bibfnamefont {H.}~\bibnamefont {Zhang}},
  \bibinfo {author} {\bibfnamefont {S.}~\bibnamefont {\ifmmode~\acute{C}\else
  \'{C}\fi{}uk}}, \bibinfo {author} {\bibfnamefont {J.}~\bibnamefont {Hu}},
  \bibinfo {author} {\bibfnamefont {M.~H.}\ \bibnamefont {Schleier-Smith}}, \
  and\ \bibinfo {author} {\bibfnamefont {V.}~\bibnamefont
  {Vuleti\ifmmode~\acute{c}\else \'{c}\fi{}}},\ }\href {\doibase
  10.1103/PhysRevA.88.063802} {\bibfield  {journal} {\bibinfo  {journal} {Phys.
  Rev. A}\ }\textbf {\bibinfo {volume} {88}},\ \bibinfo {pages} {063802}
  (\bibinfo {year} {2013})}\BibitemShut {NoStop}%
\bibitem [{\citenamefont {Haas}\ \emph {et~al.}(2014)\citenamefont {Haas},
  \citenamefont {Volz}, \citenamefont {Gehr}, \citenamefont {Reichel},\ and\
  \citenamefont {Esteve}}]{reichel40atoms}%
  \BibitemOpen
  \bibfield  {author} {\bibinfo {author} {\bibfnamefont {F.}~\bibnamefont
  {Haas}}, \bibinfo {author} {\bibfnamefont {J.}~\bibnamefont {Volz}}, \bibinfo
  {author} {\bibfnamefont {R.}~\bibnamefont {Gehr}}, \bibinfo {author}
  {\bibfnamefont {J.}~\bibnamefont {Reichel}}, \ and\ \bibinfo {author}
  {\bibfnamefont {J.}~\bibnamefont {Esteve}},\ }\href {\doibase
  10.1126/science.1248905} {\bibfield  {journal} {\bibinfo  {journal}
  {Science}\ }\textbf {\bibinfo {volume} {344}},\ \bibinfo {pages} {180}
  (\bibinfo {year} {2014})}\BibitemShut {NoStop}%
\bibitem [{\citenamefont {Dowling}\ \emph {et~al.}(1994)\citenamefont
  {Dowling}, \citenamefont {Agarwal},\ and\ \citenamefont
  {Schleich}}]{PhysRevA.49.4101}%
  \BibitemOpen
  \bibfield  {author} {\bibinfo {author} {\bibfnamefont {J.~P.}\ \bibnamefont
  {Dowling}}, \bibinfo {author} {\bibfnamefont {G.~S.}\ \bibnamefont
  {Agarwal}}, \ and\ \bibinfo {author} {\bibfnamefont {W.~P.}\ \bibnamefont
  {Schleich}},\ }\href {\doibase 10.1103/PhysRevA.49.4101} {\bibfield
  {journal} {\bibinfo  {journal} {Phys. Rev. A}\ }\textbf {\bibinfo {volume}
  {49}},\ \bibinfo {pages} {4101} (\bibinfo {year} {1994})}\BibitemShut
  {NoStop}%
\bibitem [{\citenamefont {Lvovsky}\ \emph {et~al.}(2001)\citenamefont
  {Lvovsky}, \citenamefont {Hansen}, \citenamefont {Aichele}, \citenamefont
  {Benson}, \citenamefont {Mlynek},\ and\ \citenamefont
  {Schiller}}]{PhysRevLett.87.050402}%
  \BibitemOpen
  \bibfield  {author} {\bibinfo {author} {\bibfnamefont {A.~I.}\ \bibnamefont
  {Lvovsky}}, \bibinfo {author} {\bibfnamefont {H.}~\bibnamefont {Hansen}},
  \bibinfo {author} {\bibfnamefont {T.}~\bibnamefont {Aichele}}, \bibinfo
  {author} {\bibfnamefont {O.}~\bibnamefont {Benson}}, \bibinfo {author}
  {\bibfnamefont {J.}~\bibnamefont {Mlynek}}, \ and\ \bibinfo {author}
  {\bibfnamefont {S.}~\bibnamefont {Schiller}},\ }\href {\doibase
  10.1103/PhysRevLett.87.050402} {\bibfield  {journal} {\bibinfo  {journal}
  {Phys. Rev. Lett.}\ }\textbf {\bibinfo {volume} {87}},\ \bibinfo {pages}
  {050402} (\bibinfo {year} {2001})}\BibitemShut {NoStop}%
\bibitem [{\citenamefont {Lvovsky}\ and\ \citenamefont
  {Raymer}(2009)}]{RevModPhys.81.299}%
  \BibitemOpen
  \bibfield  {author} {\bibinfo {author} {\bibfnamefont {A.~I.}\ \bibnamefont
  {Lvovsky}}\ and\ \bibinfo {author} {\bibfnamefont {M.~G.}\ \bibnamefont
  {Raymer}},\ }\href {\doibase 10.1103/RevModPhys.81.299} {\bibfield  {journal}
  {\bibinfo  {journal} {Rev. Mod. Phys.}\ }\textbf {\bibinfo {volume} {81}},\
  \bibinfo {pages} {299} (\bibinfo {year} {2009})}\BibitemShut {NoStop}%
\end{thebibliography}%

\end{document}